\UseRawInputEncoding

\documentclass[aps,showpacs,prd,twocolumn,nofootinbib,nobibnotes,]{revtex4-1}
\usepackage{graphicx}
\usepackage{amssymb}
\usepackage{amsmath}
\usepackage{color}
\usepackage{float}
\usepackage{ulem}
\usepackage{accents}
\usepackage{graphicx}
\usepackage{graphicx}
\usepackage{amsfonts}
\usepackage[colorlinks=true,
pdfstartview=FitV,linkcolor=blue,
citecolor=blue,urlcolor=blue,breaklinks=true]{hyperref}
\usepackage{array}
\usepackage{float}
\usepackage{placeins}
\usepackage[dvipsnames]{xcolor}
\usepackage{csquotes}
\usepackage{bbold}
\usepackage{units}
\usepackage{fixmath}

\newcolumntype{C}[1]{>{\centering\arraybackslash}m{#1}}
\renewcommand{\eqref}[1]{\mbox{Eq.~(\ref{#1})}}

\definecolor{ForestForestGreen}{rgb}{0.13,0.55,0.13}

\begin{document}

\title{Rotatory power reversal induced by magnetic-current in bi-isotropic media}
\author{Pedro D. S. Silva$^a$}
\email{pedro.dss@discente.ufma.br} \email{pdiego.10@hotmail.com}
\author{Manoel M. Ferreira Jr.$^a$}
\email{manojr.ufma@gmail.com}
\affiliation{$^a$Departamento de F\'{\i}sica, Universidade Federal do Maranh\~{a}o,
Campus Universit\'{a}rio do Bacanga, S\~{a}o Lu\'is (MA), 65080-805, Brazil}

\begin{abstract}
Bi-isotropic media constitute a proper scenario for scrutinizing optical effects stemming from magnetoelectric parameters. Chiral magnetic current is a macroscopic effect arising from the chiral magnetic effect that enriches the phenomenology of a classical dielectric medium. This work examines optical aspects of bi-isotropic media in the presence of magnetic conductivity. The full isotropic scenario manifests circular birefringence described by a dispersive rotatory power that changes sign at a given frequency. For a bi-isotropic medium with antisymmetric magnetic conductivity, an intricate dispersive rotatory power is attained, supplied with sign reversal as well. This scenario also indicates a handedness reversion of the medium, an unusual property in dielectrics, which may work as a signature of bi-isotropic media supporting chiral magnetic current.
\end{abstract}

\maketitle
\affiliation{$^a$Departamento de F\'{\i}sica, Universidade Federal do Maranh\~{a}o,
Campus Universit\'{a}rio do Bacanga, S\~{a}o Lu\'is (MA), 65080-805, Brazil}
\affiliation{$^a$Departamento de F\'{\i}sica, Universidade Federal do Maranh\~{a}o,
Campus Universit\'{a}rio do Bacanga, S\~{a}o Lu\'is (MA), 65080-805, Brazil}
\affiliation{$^a$Departamento de F\'{\i}sica, Universidade Federal do Maranh\~{a}o,
Campus Universit\'{a}rio do Bacanga, S\~{a}o Lu\'is (MA), 65080-805, Brazil}

\section{\label{section1}Introduction}

The classical propagation of electromagnetic waves in continuous matter is ruled by the Maxwell equations and the constitutive relations that enclose the response to applied electromagnetic fields \cite{Jackson, Zangwill}. In linear electrodynamics, a great diversity of electromagnetic phenomena in continuous media can be accounted for by general constitutive relations,
\begin{subequations}
\label{magnetic-biisotropic-1}
\begin{align}
D^{i} &= \epsilon_{ij} E^{j} + \alpha_{ij} B^{j} , \label{magnetic-biisotropic-1-1} \\
H^{i} &= \mu^{-1}_{ij}B^{j} + \beta_{ij} E^{j}, \label{magnetic-biisotropic-1-2}
\end{align}
\end{subequations}
where $\epsilon_{ij}$ and $\mu_{ij}$ are the electric permittivity and magnetic permeability tensors, while $\alpha_{ij}$ and $\beta_{ij}$ are magnetoelectric parameters that capture the electric (magnetic) response to applied magnetic (electric) field, respectivelly. In simplest scenario, the relations (\ref{magnetic-biisotropic-1}) read
\begin{align}
\begin{pmatrix}
{\bf{D}}\\
{\bf{H}}
\end{pmatrix} &= \begin{pmatrix}
\epsilon & \alpha \\
\beta & \mu^{-1} 
\end{pmatrix}  \begin{pmatrix} 
{\bf{E}} \\
{\bf{B}}
\end{pmatrix}, \label{magnetic-biisotropic-2}
\end{align}
where $\epsilon$, $\alpha$, $\beta$, $\mu$ are parameters composing the bi-isotropic linear connection between $(\mathbf{D},\mathbf{H%
})$ and $(\mathbf{E},\mathbf{B})$ for homogenous non-dispersive materialss \cite{Sihvola, Kong}.

In the last decades, bi-isotropic relations have been extensively investigated \cite{Sihvola1, Sihvola2, Nieves, Gauthier, Aladadi}, being also important to address properties of topological insulators \cite{Chang1, Urrutia, Urrutia2, Lakhtakia, Winder, Li,Li1} and axion electrodynamics \cite{Sekine, Tobar, BorgesAxion}. In nonlinear electrodynamics, the constitutive tensors usually depend on the electric and magnetic fields, providing a way to describe well-known phenomena, e. g., the Kerr and Cotton-Mouton effects \cite{Lorenci}. Most recently, nonlinear electrodynamics addressing axion-photons couplings has been reported \cite{Helayel1}, where general nonlinear permittivity and permeability tensors present dependence on the frequency, wave vector, and axion mass. Other aspects of nonlinearities in constitutive tensors, such as the description of ferrimagnetic materials and radiation produced by a charged particle interacting with a nonlinear medium have also been examined \cite{Helayel2}.

For scenarios where the constitutive parameters in \eqref{magnetic-biisotropic-1} are complex tensors, they characterize bi-anisotropic media \cite{Yakov,Damaskos}, also known as chiral materials \cite{Lin, Sihvola3, Pedro3}. A medium is named chiral when it lacks inversion symmetry, that is, when it is parity-odd, or, in other words, when it is invariant only under the $L_0$ component of the Lorentz group \cite{Hillion}. A chiral scenario, realized by constitutive relations (\ref{magnetic-biisotropic-2}) with complex parameters, was used to examine the Fresnel equations that describe reflection, refraction, and Brewster angle \cite{Hillion}. Right-handed (RCP) and left-handed circularly polarized (LCP) waves are examples of electromagnetic chiral phenomena, which in an achiral medium propagate at the same phase velocity, yielding no anisotropy effect.  In a chiral medium, however, the LCP and RCP waves travel at distinct phase velocities (related to different refractive indices), yielding the rotatory power (RP) that measures the birefringence (optical rotation) \cite{Kaushik2, Barron2, Hecht, Wagniere,TangPRL}. It is known that the optical rotation stems from the natural optical activity of the medium or can be induced by external fields, as it occurs in the Faraday effect \cite{Fowles, Condon, Bennett}. Magneto-optical effects constitute a useful probe to investigate topological insulators \cite{Ohnoutek, Tse, Maciejko} and new materials \cite{Crasee,Shimano}.
	
When the optical rotation depends on the frequency, there occurs RP dispersion. If it depends on the frequency and undergoes reversion, it is called anomalous RP dispersion \cite{Newnham, Tschugaeff}. Thus, the RP analysis is a relevant tool to describe optical properties of several systems, such as crystals \cite{Dimitriu, Birefringence1}, organic compounds \cite{Barron2, Xing-Liu}, rotating plasma and neutron stars \cite{Gueroult, Gueroult2}, graphene terahertz magneto-optical phenomena \cite{Poumirol}, gas of fast-spinning molecules \cite{Tutunnikov, Steinitz}, entangled photons in quantum optics \cite{Tischler}, and chiral metamaterials \cite{Woo, Zhang}. In addition to birefringence, circular dichroism, the difference in the absorption of LCP and RCP light, is also used to probe the optical activity of materials. Dichroism can work as a tool to distinguish between Dirac and Weyl semimetals \cite{Hosur}, perform enantiomeric discrimination \cite{Nieto-Vesperinas,Tang}, and for developing graphene-based devices at terahertz frequencies \cite{Amin}. 

Another phenomenon that has caught attention in the context of continuous media is the Chiral Magnetic Effect (CME), which involves a macroscopic generation of an electric current in the presence of a magnetic field as the result of an asymmetry between the number density of left- and right-handed chiral fermions \cite{Kharzeev1,Kharzeev1B,Fukushima,Gabriele, Bubnov}. In condensed-matter systems, the CME plays a relevant role in Weyl semimetals, where it is usually connected to the chiral anomaly associated with Weyl nodal points \cite{Burkov}, but also takes place in the absence of Weyl nodes \cite{Chang}. Other investigations on this subject discuss, for instance, the anisotropic effects stemming from tilted Weyl cones \cite{Wurff}, the CME and anomalous transport in Weyl semimetals \cite{Landsteiner}, quantum oscillations arising from the CME \cite{Kaushik}, the determination of electromagnetic fields produced by an electric charge near a topological Weyl semimetal with two Weyl nodes \cite{Ruiz}. In an effective approach, the magnetic current can be introduced using a general constitutive relation for the current density of the form $J^{i}=\sigma E^{i}+\sigma^{B}_{ij} B^{j}$, where $\sigma$ and $\sigma^{B}$ are the Ohmic conductivity and magnetic-conductivity, respectively. A dielectric medium supporting such a magnetic current has been recently examined for symmetric and antisymmetric conductivity tensors \cite{Pedro1}, with the determination of the refractive indices, propagating modes, and some optical properties. The antisymmetric scenario for $\sigma^{B}$ has found realization in some Weyl semimetals \cite{Kaushik1}. The interplay between Lorentz-violating electrodynamics, optical effects, and condensed matter physics \cite{Pedro2, Marco} has also been explored.

Considering the relevance of optical effects for describing material properties, in this work we investigate the propagation of electromagnetic waves in bi-isotropic media endowed with magnetic conductivity, a type of chiral scenario not yet examined. We firstly address a bi-isotropic dielectric with an isotropic current constitutive relation, focusing on the birefringence and rotatory power. We obtain a linearly dispersive RP with sign reversion, an effect not reported in dielectrics before. At a second moment, we develop the same analysis for an antisymmetric conductivity tensor, attaining an involved dispersive RP which also exhibits sign change. These results may also cause handedness reversal of the medium. This paper is outlined as follows: in Sec.~\ref{section-dispersion-relation} we present the general framework to obtain the dispersion relations, refractive indices, and propagating modes. In Sec.~\ref{section-dispersion-relations-particular-cases}, we address a bi-isotropic medium in the presence of isotropic and antisymmetric magnetic conductivity and examine the results. Finally, we summarize our results in Sec.~\ref{final-remarks}.

\section{\label{section-dispersion-relation} Dispersion relation for
bi-isotropic medium endowed with $\sigma^{B}$}

We take as starting point the Maxwell equations for a homogeneous and linear medium, here written in accordance with a plane wave ansatz, 
\begin{subequations}
	\begin{align}
	\mathbf{k}\cdot \mathbf{D}& =0\,,\quad \mathbf{k}\times \mathbf{H}+\omega
	\mathbf{D}=-{\mathrm{i}\mathbf{J}, }      \label{maxwell-nonhomogeneous-1}           \\[1ex] 
	\mathbf{k}\cdot \mathbf{B}& =0\,,\quad  \omega \mathbf{B}=\mathbf{k}\times
		\mathbf{E}\,.
	\end{align}
\end{subequations}
We now suppose the validity of isotropic-anisotropic constitutive relations, in which the anisotropy is constrained into the magnetoelectric parameters, that is, the electric permittivity and magnetic permeability are written as
\begin{equation}
\zeta _{ka}={\mu }^{-1}\delta _{ka},\text{ \ \ }\epsilon _{ij}=\epsilon
\delta _{ij},  \label{ex3}
\end{equation}
so that the constitutive relations (\ref{magnetic-biisotropic-1}) take the form,
\begin{subequations}
	\label{constitutive-relations-general-2}
	\begin{eqnarray}
	D^{i} &=&\epsilon \delta_{ij}E^{j}+\alpha _{ij}B^{j},  \label{DEB4a} \\
	H^{i} &=&\beta _{ij}E^{j}+\mu^{-1}\delta_{ij}B^{j}.  \label{HEB4a}
	\end{eqnarray}
\end{subequations}
For obtaining wave equations that describe the propagation of electromagnetic waves in such a dielectric medium, we replace the constitutive relation (\ref{DEB4a}) in the second relation of \eqref{maxwell-nonhomogeneous-1} and use Faraday's law, $\omega \mathbf{B}=\mathbf{k\times E}$, writing
\begin{align}
\frac{1}{\mu\omega }\left[ \mathbf{k}\times \left(\mathbf{k}\times \mathbf{E}\right)\right]
^{i}& +\omega \epsilon \delta_{ij}E^{j}+\beta
_{ka}\epsilon _{ijk}k^{j}E^{a}+  \notag \\
& +\alpha _{ij}\epsilon _{jmn}k^{m}E^{n}=-\mathrm{i}\sigma^{B}_{ij} B^{j},
\label{ex2}
\end{align}%
where we have supposed the constitutive relation for the magnetic current density,
 \begin{equation}
 J^{i}=\sigma^{B}_{ij} B^{j}. 
 \end{equation}
with a general magnetic-conductivity tensor, $\sigma^{B}_{ij}$. Using again the Faraday's law, \eqref{ex2} is rewritten as
\begin{equation}
\left[ \mathbf{k}\times \left(\mathbf{k}\times \mathbf{E}\right)\right]
^{i}+\omega ^{2}\mu \bar{\epsilon}_{ij}E^{j}=0,  \label{ex8}
\end{equation}%
where
\begin{equation}
\bar{\epsilon}_{in}(\omega )=\epsilon \delta _{in}+\frac{1}{\omega }  \left( \beta
_{kn}\epsilon _{imk}+\alpha _{ij}\epsilon _{jmn}+ \mathrm{i} \frac{\sigma^{B}_{ij}}{\omega} \epsilon_{jmn}\right) k^{m},  \label{ex7}
\end{equation}%
defines the frequency-dependent extended permittivity tensor, which carries
the electric and magnetic response of the medium.  \eqref{ex8} can be cast in the form,
\begin{equation}
M_{ij}E^{j}=0,  \label{ex11}
\end{equation}
where the tensor $M_{ij}$ is given in terms of the refractive index, $\mathbf{n}$, as
\begin{align}
M_{ij}&=n^{2}{\delta }_{ij}-n_{i}n_{j}-c^{2}\mu \bar{{\epsilon }}_{ij} .
\label{ex12} 
\end{align}
We have used $\mathbf{k}=\omega
\mathbf{n}/c$, with $c$ being the vacuum light speed. 
Here we consider that the index $n$ is nonnegative, so that we adopt $n=+\sqrt{\mathbf{n}^{2}}$, instead of $|\mathbf{n}|$, in order to permit complex refractive indices. The refractive indices with negative real parts, related to metamaterials, are not considered here. To find the nontrivial solution for the electric field of \eqref{ex11}, we impose the condition $\mathrm{det}[M_{ij}]=0$ in order to attain the dispersion relations
that govern the wave propagation in the medium. 


\section{\label{section-dispersion-relations-particular-cases}Dispersion relations, refractive indices and propagating modes}

In this section, we examine the propagation of electromagnetic waves in a
dielectric medium under the validity of the relations (\ref{ex8}) and (\ref{ex7}) for two configurations of the magnetic conductivity: the symmetric isotropic one and the antisymmetric one.

\subsection{\label{full-isotropic-case}Full-isotropic case \label{Full-isotropic-case}}

In the context of the constitutive relations (\ref{constitutive-relations-general-2}), we begin considering the total
symmetric isotropic configuration for the quantities $\alpha_{ij}$,
$\beta_{ij}$ and $\sigma^{B}_{ij}$, in which these tensor are written as diagonal tensors, namely
\begin{equation}
\alpha _{ij} =\alpha \delta _{ij}, \quad \beta _{ij} =\beta \delta _{ij} , \quad \sigma^{B}_{ij}=\Sigma \delta_{ij}, 
\label{isot5}
\end{equation}
with $\alpha$, $\beta \in \mathbb{C}$, and $\Sigma \in \mathbb{R}$. The condition $\alpha_{ij}=-\beta_{ij}^{\dagger}$ for electromagnetic energy conservation \cite{Pedro3}, when applied on the parametrization of \eqref{isot5}, yields
\begin{align}
\beta^{*}=-\alpha. \label{isotropic-case-1-1}
\end{align}
In this case, the constitutive relations take on the typical bi-isotropic form,
\begin{subequations}
\label{constitutive-relations-biisotropic-1}
\begin{align}
\mathbf{D}& =\epsilon \mathbf{E}+\alpha \mathbf{B},
\label{constitutive-iso1D} \\
\mathbf{H}& =\frac{1}{\mu }\mathbf{B}+\beta \mathbf{E},
\label{constitutive-iso2D} \\
\mathbf{J}& ={\Sigma}\mathbf{B},
\label{constitutive-iso3D}
\end{align}%
\end{subequations}
being the last one the isotropic magnetic current. As already mentioned, such relations play
relevant role in topological insulators \cite{Chang1, Urrutia, Urrutia2,
Lakhtakia, Winder, Li,Li1} and axion systems \cite{Sekine, Tobar,
BorgesAxion}.

Inserting relations (\ref{isot5}) in \eqref{ex7}, one obtains
\begin{equation}
c \bar{\epsilon}_{ij}=c \epsilon  \delta _{ij}-\left(\alpha +\beta +\mathrm{i} \frac{\Sigma}{\omega} \right) \epsilon _{ijm}n^{m},
\label{iso4}
\end{equation} 
where the last term in RHS represents the ``magnetic-electric" contribution to the medium permittivity. In
this case, the tensor $M_{ij}$ (\ref{ex12}) has the form
\begin{subequations}
\label{m-matrix-full-isotropic-0}
\begin{align}
[M_{ij}] = \mathcal{M} + \mu c\left( \alpha+\beta+\mathrm{i}\frac{\Sigma}{\omega}\right) \mathcal{S}   , \label{m-matrix-full-isotropic-1}
\end{align}
where 
\label{compact-M-matrix-full-isotropic-case-1}
\begin{align}
\mathcal{M} &= \left(n^{2} -c^{2}\mu \epsilon \right) \mathbb{1}_{3} - \begin{pmatrix}
n^{2}_{1} & n_{1}n_{2} & n_{1}n_{3} \\
n_{1}n_{2} & n_{2}^{2} & n_{2} n_{3} \\
n_{1}n_{3} & n_{2}n_{3} & n_{3}^{2}
\end{pmatrix} , \label{compact-M-matrix-full-isotropic-case-2}
\end{align}
and
\begin{align}
\mathcal{S}&= \begin{pmatrix}
0& - n_{3} & n_{2} \\
n_{3} & 0 & - n_{1} \\
-n_{2} &  n_{1} & 0
\end{pmatrix} . \label{compact-M-matrix-full-isotropic-case-3}
\end{align}
\end{subequations}

Requiring $\mathrm{det}[M_{ij}]=0$, one gets
\begin{equation}
n^{4}-n^{2} c^{2}\left[ 2\mu \epsilon  -\mu ^{2} \left(\alpha +\beta+\mathrm{i}\frac{\Sigma}{\omega} \right)^{2}\right] +\mu ^{2} \epsilon^{2}c^{4} =0  .  \label{iso6}
\end{equation}%
Solving for $n$, we obtain the following refractive indices
\begin{equation}
n_{\pm }=c\sqrt{\mu {\epsilon} -Z} \pm \mathrm{i}c\sqrt{Z}  ,
\label{iso10}
\end{equation}
with
\begin{align}
Z&=\frac{\mu ^{2}}{4} \left(\alpha +\beta+\mathrm{i}\frac{\Sigma}{\omega} \right)^{2} .  \label{iso-7-1}
\end{align}

As we have started with isotropic tensors, $\epsilon \delta
_{ij} $, $\mu ^{-1}\delta _{ij}$, $\alpha \delta _{ij}$, $\beta \delta _{ij}$, any arising anisotropy effects stem from the way magnetoelectric and magnetic conductivity are coupled to the fields in the constitutive relations (\ref{constitutive-relations-biisotropic-1}).

\subsubsection{Propagation modes} 

In order to achieve the propagating modes, we write the refractive indices (\ref{iso10}) as
\begin{align}
n_{\pm}^{2} = \mu\epsilon \pm 2 \mathrm{i}c \sqrt{Z} n_{\pm}, \label{full-iso-propagation-1}
\end{align} 
and replace it in the matrix (\ref{m-matrix-full-isotropic-0}), so that \eqref{ex11} yields
\begin{align}
{\bf{E}}_{\pm} &= \frac{1}{\sqrt{2}n\sqrt{n^{2}-n_{1}^{2}}} \begin{pmatrix}
n^{2}-n_{1}^{2} \\
\pm \mathrm{i} n n_{3} -n_{1}n_{2} \\
\mp \mathrm{i} n n_{2} - n_{1} n_{3} 
\end{pmatrix} . \label{full-iso-propagation-2}
\end{align}

Now, we consider the specific case of propagation along the $z$-axis, ${\bf{n}}=(0,0,n)$, in such a way \eqref{full-iso-propagation-2} provides right-handed ($\hat{\bf{E}}_{-}$) and left-handed circular ($\hat{\bf{E}}_{+}$) polarization vectors, respectively,
 \cite{Jackson, Zangwill},
\begin{align}
{\bf{E}}_{\pm} &=\frac{1}{\sqrt{2}} \begin{pmatrix}
1\\
\pm \mathrm{i} \\
0
\end{pmatrix} . \label{full-iso-propagation-3}
\end{align} 
As well known, the circular polarization modes (\ref{full-iso-propagation-3}) open the possibility for the attainment of circular birefringence in this system.

\subsubsection{Optical effects of complex magnetoelectric parameters in
non-conducting dielectrics}

As the modes (\ref{full-iso-propagation-3}) do
not depend on the real or complex character of the parameters $\alpha$ and $\beta$, we will examine both situations. In this sense, we suppose $\alpha$ and $\beta$  as complex parameters,
\begin{equation}
\alpha=\alpha' +\mathrm{i} \alpha'', \quad \beta=\beta' +\mathrm{i} \beta'',
\end{equation}
where $\alpha'=\mathrm{Re}[\alpha]$, $\alpha''=\mathrm{Im}[\alpha]$, $\beta'=\mathrm{Re}[\beta]$ and $\beta''=\mathrm{Im}[\beta]$. The condition (\ref{isotropic-case-1-1}) then implies
\begin{equation}
\alpha'=-\beta', \quad \alpha''=\beta'', \label{condition-magnetoelectric-parameters-1}
\end{equation} 
so that 
\begin{equation}
\alpha+\beta= 2{\mathrm{i}\alpha''}.
\label{alpha+beta1}
\end{equation} 
In order to examine the physical behavior
	stemming from the constitutive relations (\ref{constitutive-iso1D}), (%
	\ref{constitutive-iso2D}) and (\ref{constitutive-iso3D}), we rewrite the refractive index (\ref{iso10}) using \eqref{alpha+beta1}, that is,
\begin{align}
n_{\pm}&=c \sqrt{\mu \epsilon + \mu^{2} \left(\alpha''+\frac{\Sigma}{2\omega}\right)^{2}} \mp \mu c \left(\alpha''+\frac{\Sigma}{2\omega}\right), \label{isotropic-case-1-2-0-1}
\end{align} 
which yields two real and positive values, implying birefringence. The isotropic magnetic-conductivity $\Sigma$ appears as a frequency-dependent contribution, which brings about a dispersive behavior to the system. Since the propagating modes are described by circularly polarized vectors, see \eqref{full-iso-propagation-3},
the birefringence effect can be evaluated in terms of the rotatory power \cite{Fowles, Hecht, Pedro2}, 
\begin{equation}
\delta =-\frac{[\mathrm{Re}(n_{+})-\mathrm{Re}(n_{-})]\omega }{2}\
\label{eq:rotatory-power1A},
\end{equation} 
which, with indices (\ref{isotropic-case-1-2-0-1}), yields
\begin{equation}
\delta = \frac{\mu c \Sigma}{2} +\mu c\omega \alpha ^{\prime \prime }. \label{isotropic-case-1-3}
\end{equation} 
It displays a zeroth order term in the frequency that recovers the same rotatory power of isotropic case of Ref. \cite{Pedro1} when $\alpha ^{\prime \prime }=0$. For a negative $\alpha ^{\prime \prime}$, the rotatory power reads
\begin{equation}
\delta = \frac{\mu c \Sigma}{2} -\mu  c \omega |\alpha ^{\prime \prime }|. \label{isotropic-case-1-4}
\end{equation} 
From \eqref{isotropic-case-1-4}, one finds a cut-off frequency $\omega'$,
\begin{align}
\omega' &=  \frac{ \Sigma}{2|\alpha''|} , \label{cutoff-frequency-full-isotropic-case}
\end{align}
defining the value at which there occurs the sign reversion of $\delta$. Considering usual RP sign convention \cite{Hecht} and \eqref{isotropic-case-1-4}, one notices:
\begin{itemize}
	\item For $0<\omega < \omega'$, one has $\delta >0$, causing a clockwise rotation of the linear electric field polarization.
	\item For $\omega > \omega'$, one finds $\delta <0$, implying a counterclockwise rotation of the linear polarization.
\end{itemize}

As this RP inversion takes place at a positive frequency, $\omega'>0$, one needs to have either $\Sigma <0$ or $\alpha''<0$. In the following we shall consider $\alpha'' <0$. The RP reversion here observed is not usual in ordinary linear dielectric nor in bi-isotropic or bi-anisotropic dielectric media. It also does not appear in media endowed with isotropic magnetic conductivity \cite{Pedro1}, see \eqref{full-isotropic-alpha-beta-nulos-1}. However, it is reported in rotating plasmas \cite{Gueroult} and graphene systems \cite{Poumirol}. Furthermore, as shown in \eqref{isotropic-case-1-4}, such an effect also appears in
a bi-isotropic dielectric under the presence of the magnetic current  (\ref{constitutive-iso3D}). 

\begin{figure}[h!]
\begin{centering}
\includegraphics[scale=0.68]{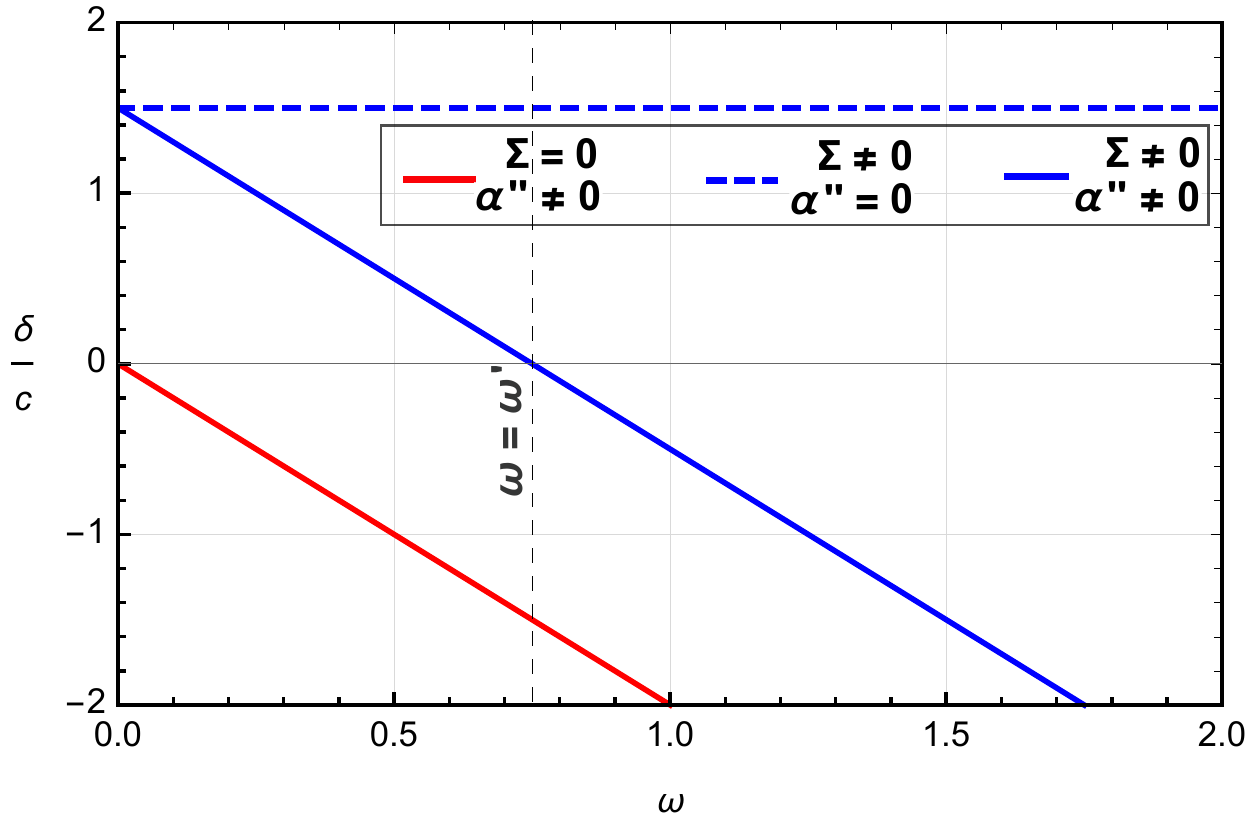}
\par\end{centering}
\caption{\label{plot-rotatory-dispersion-fullisotropic-case} Polarization rotation angle per length unit, as given in \eqref{isotropic-case-1-4}. Here, we have used $\mu=1$, $\Sigma=3$ and $|\alpha''|=2$. The vertical dashed line is given by $\omega' = 3/4$, with $\omega'$ of \eqref{cutoff-frequency-full-isotropic-case}.}
\end{figure}

 The general behavior of the rotatory power (\ref{isotropic-case-1-4}) as function of frequency is illustrated in Fig. \ref{plot-rotatory-dispersion-fullisotropic-case} for some values of $\Sigma$, $\alpha''$ and $\mu$.  
Finally, for $\omega=\omega'$, the birefringence does not occur ($\delta=0$).

Note that for $\alpha$, $\beta \in \mathbb{R}$, one has simply $\beta=-\alpha$ and $\beta+\alpha=0$, since $\alpha''=0$. Hence, \eqref{isotropic-case-1-2-0-1} becomes
\begin{align}
n_{\pm} &=c \sqrt{\mu\epsilon + \frac{\mu^{2} \Sigma^{2}}{4\omega^{2}}} \mp \frac{\mu c \Sigma}{2\omega}, \label{full-isotropic-alpha-beta-nulos-1}
\end{align} 
which also leads to birefringence, in this case stemming entirely from the magnetic conductivity. The corresponding rotatory power, $\delta=\mu c\Sigma/2$, obviously does not undergo reversion.

\subsection{\label{bi-isotropic-anti-symmetric}Bi-isotropic case with off-diagonal antisymmetric magnetic-conductivity tensor}

Let us now consider a substrate described by the bi-isotropic constitutive parameters in the presence of a chiral current written in terms of an antisymmetric magnetic conductivity, that is, 
\begin{align}
\alpha_{ij} =\alpha \delta_{ij}, \quad \beta_{ij} = \beta \delta_{ij}, \quad \sigma^{B}_{ij} = \epsilon_{ijk}b_{k} , \label{anti-symmetic-constitutive-parameters}
\end{align} 
for which the permittivity (\ref{ex7}) is
\begin{align}
c \bar{\epsilon}_{ij}&=  \left[c \epsilon - \frac{ \mathrm{i}}{\omega} ({\bf{b}}\cdot {\bf{n}}) \right] \delta_{ij} + (\alpha+\beta) \epsilon_{ijk}n_{k} + \frac{\mathrm{i}}{\omega} n_{i}b_{j}. \label{anti-symmetric-case-1}
\end{align}  
The matrix $[M_{ij}]$ of \eqref{ex11} reads
\begin{align}
[M_{ij}] &= \mathcal{M} +\frac{ \mathrm{i} \mu c }{\omega} ({\bf{b}}\cdot {\bf{n}}) \mathbb{1}_{3} -\frac{\mathrm{i}\mu c}{\omega}\mathcal{B} ,  \label{anti-symmetric-case-2}
\end{align} 
with $\mathcal{M}$ given by \eqref{compact-M-matrix-full-isotropic-case-2} and 
\begin{subequations}
\label{anti-symmetric-case-3-0}
\begin{align}
\mathcal{B} &= \begin{pmatrix}
n_{1} b_{1} & - \mathrm{i}\omega \gamma n_{3}+ n_{1} b_{2} & \mathrm{i}\omega \gamma n_{2} + n_{1} b_{3} \\
\mathrm{i}\omega \gamma n_{3} + n_{2}b_{1} & n_{2} b_{2} & -\mathrm{i}\omega \gamma n_{1} + n_{2} b_{3} \\
-\mathrm{i}\omega \gamma + n_{3}b_{1} & \mathrm{i}\omega \gamma n_{1} + n_{3}b_{2} & n_{3} b_{3}
\end{pmatrix} , \label{anti-symmetric-case-3-1}
\end{align}
\end{subequations}
and $\gamma = \alpha+\beta.$ Evaluating $\mathrm{det}[M_{ij}]=0$, one attains the following dispersion equation:
\begin{align}
0&=n^{4} - 2 n^{2} \left[\mu\epsilon c^{2}- \frac{\mu^{2} c^{2}}{2} (\alpha+\beta)^{2} -\frac{\mathrm{i}\mu c}{\omega} ({\bf{b}}\cdot {\bf{n}}) \right] + \nonumber \\
&\phantom{=}+ \mu^{2}c^{2} \left[c \epsilon -\frac{\mathrm{i}}{\omega} ({\bf{b}}\cdot {\bf{n}})\right]^{2} =0 . \label{anti-symmetric-case-4}
\end{align} 
Implementing ${\bf{b}}\cdot {\bf{n}} = b n \cos\theta$ in \eqref{anti-symmetric-case-4}, we search for the solutions for $n$ that recover the refractive indices of an ordinary dielectric, $n_{\pm} \mapsto c \sqrt{\mu\epsilon}$, in the limit of vanishing magnetoelectric and conductivity parameters, that is, $\gamma \mapsto 0$ and $b\mapsto 0$. We thus find,
\begin{align}
n_{\pm}&= c \sqrt{\mu\epsilon - \frac{\mu^{2}}{4} \left(\gamma \mp \frac{b\cos\theta}{\omega}\right)^{2} } \pm \frac{\mathrm{i}\mu c}{2} \left(\gamma\mp \frac{b\cos\theta}{\omega} \right) . \label{anti-symmetric-case-5}
\end{align}
The other two solutions of \eqref{anti-symmetric-case-4} that provide $n_{\pm} \mapsto - c\sqrt{\mu\epsilon}$ in the limit of ordinary dieletric will not be considered here. Considering the relations  (\ref{condition-magnetoelectric-parameters-1}), the refractive indices are rewritten now as
\begin{align}
n_{\pm}&= c\sqrt{\mu\epsilon -\frac{\mu^{2}}{4} \left(2\mathrm{i}\alpha'' \mp \frac{b\cos\theta}{\omega}\right)^{2}} \mp \mu c \alpha'' - \frac{\mathrm{i}\mu c}{2\omega}b\cos\theta , \label{anti-symmetric-case-6}
\end{align} 
which can be expressed separating the real and complex pieces,
\begin{subequations}
\label{anti-symmetric-case-7-0}
\begin{align}
n_{\pm} = U A_{+} \mp \mu c \alpha'' + \mathrm{i} \left( \pm U A_{-} - \frac{\mu c}{2\omega} b\cos\theta\right) , \label{anti-symmetric-case-7}
\end{align}
with
\begin{align}
U &= \frac{c}{\sqrt{2}} \sqrt{\mu\epsilon + \mu^{2} \alpha''^{2} - \frac{\mu^{2}}{4\omega^{2}} b^{2} \cos^{2}\theta } , \label{anti-symmetric-case-8} \\
A_{\pm} &= \sqrt{\sqrt{ 1 + \left( \frac{\mu^{2} \alpha'' b\cos\theta}{ \omega (\mu\epsilon + \mu^{2} \alpha''^{2}) - \frac{\mu^{2} b^{2}}{4\omega} \cos^{2}\theta } \right)^{2} } \pm 1 }  . \label{anti-symmetric-case-9}
\end{align}
\end{subequations}

We point out that there is a frequency window at which $U$ is purely imaginary, i. e., for $0<\omega < \omega_{0}$, where $\omega_{0}$ is 
\begin{align}
\omega_{0} = \frac{1}{2} \sqrt{\frac{\mu b^{2}\cos^{2}\theta}{\epsilon + \mu \alpha''^{2}}}. \label{anti-symmetric-case-extra-1}
\end{align}
In this range, the refractive indices are given by
\begin{align}
n_{\pm} &= \mp U' A_{-} \mp \mu c \alpha'' + \mathrm{i} \left( U' A_{+} - \frac{\mu c}{2\omega} b\cos\theta \right) , \label{anti-symmetric-case-extra-2}
\end{align}
with
\begin{align}
U' &=\frac{c}{\sqrt{2}} \sqrt{\frac{\mu^{2}}{4\omega^{2}} b^{2}\cos^{2}\theta - \mu \epsilon - \mu^{2}\alpha''^{2}} . \label{anti-symmetric-case-extra-3}
\end{align}

Therefore, two main scenarios arise: i) for $0< \omega <\omega_{0}$, the refractive indices are given by \eqref{anti-symmetric-case-extra-2}; and ii) for $\omega > \omega_{0}$, they are described by \eqref{anti-symmetric-case-7}. The physical consequences of this feature will be examined later, with a focus on the rotatory power.

\subsubsection{Propagation modes}

Let us consider the coordinate system where propagation is along the $z$-axis, ${\bf{n}}=(0, 0 , n)$, then \eqref{anti-symmetric-case-2} simplifies as
\begin{align}
[M_{ij}] &= (n^{2}-c^{2}\mu\epsilon)\mathbb{1}_{3} + \nonumber \\
&\phantom{=}+ c \begin{pmatrix}
 +\frac{ \mathrm{i}\mu n b}{\omega} \cos\theta & - 2\mathrm{i}\mu \alpha'' n & 0 \\
2 \mathrm{i}\mu \alpha'' n & +\frac{ \mathrm{i}\mu n b}{\omega} \cos\theta   & 0 \\
-\mathrm{i}\mu b_{1}n /\omega & -\mathrm{i}\mu b_{2} n / \omega  & -n^{2}/c
\end{pmatrix} . \label{anti-symmetric-case-12}
\end{align} 
Rewritting \eqref{anti-symmetric-case-6} as
\begin{align}
n_{\pm}^{2}=c^{2}\mu\epsilon \pm \mathrm{i} \mu c \left(2\mathrm{i} \alpha'' \mp \frac{b}{\omega}\cos\theta\right) n_{\pm} . \label{anti-symmetric-case-13}
\end{align} 
and replacing it into \eqref{anti-symmetric-case-12}, the condition (\ref{ex11}) provides the following electric fields for the propagating modes:
\begin{align}
{\bf{E}}_{\pm} &= E_{0} \begin{pmatrix}
1\\
\pm \mathrm{i} \\
-\mathrm{i} n_{\pm}(b_{1}\pm \mathrm{i}b_{2})  / ( \omega \epsilon c) 
\end{pmatrix} ,\label{anti-symmetric-case-14}
\end{align} 
with an appropriately chosen amplitude, $E_{0}$. Such fields stand for to ``mixed modes'', composed by circular polarization transversal sector and an additional longitudinal component. 

For the special case where the ${\bf{b}}$ vector is parallel or anti-parallel to the propagation direction ${\bf{n}}$, we set ${\bf{b}}=(0, 0, b_{3})$, with which \eqref{anti-symmetric-case-14} yields
\begin{align}
{\bf{E}}_{\pm} &=E_{0} \begin{pmatrix}
1\\
\pm \mathrm{i} \\
0
\end{pmatrix} , \label{anti-symmetric-case-15}
\end{align}
representing left- and right-handed circularly polarized vectors, respectively.

\subsubsection{Optical effects}

Considering the circular polarization of the propagating modes (\ref{anti-symmetric-case-15}), the rotatory power (\ref{eq:rotatory-power1A}) represents properly the birefringence effects. Thus, for the refractive indices given in \eqref{anti-symmetric-case-7-0}, one achieves
\begin{align}
\delta &= \mu c \omega \alpha'' ,\label{anti-symmetric-case-16}
\end{align} 
defined for the regime $\omega >\omega_{0}$, where $U$ is real.
Differently from the previous scenario of Sec. \ref{full-isotropic-case}, the rotatory power  (\ref{anti-symmetric-case-16}) can not undergo sign reversion. Furthermore, the result in \eqref{anti-symmetric-case-16} recovers the same result obtained in Ref. \cite{Pedro3}, so that the magnetic conductivity does not contribute in this frequency range.

On the other hand, for the frequency window $0 <\omega <\omega_{0}$,   the refractive indices are the ones of \eqref{anti-symmetric-case-extra-2}, yielding the following rotatory:
\begin{align}
\delta &= \frac{c \omega A_{-}} {\sqrt{2}} \sqrt{ \frac{\mu^{2}}{4\omega^{2}} b^{2}\cos^{2} \theta-\mu\epsilon - \mu^{2} \alpha''^{2}} + \mu \omega c \alpha'' . \label{anti-symmetric-case-rotatory-extra-1}
\end{align}
It is worthy to mention that $\delta>0$ for $\alpha''<0$ or $\alpha''>0$ and that \eqref{anti-symmetric-case-rotatory-extra-1} holds only in the frequency range $0 <\omega <\omega_{0}$.

\begin{figure}[h]
\begin{centering}
\includegraphics[scale=0.68]{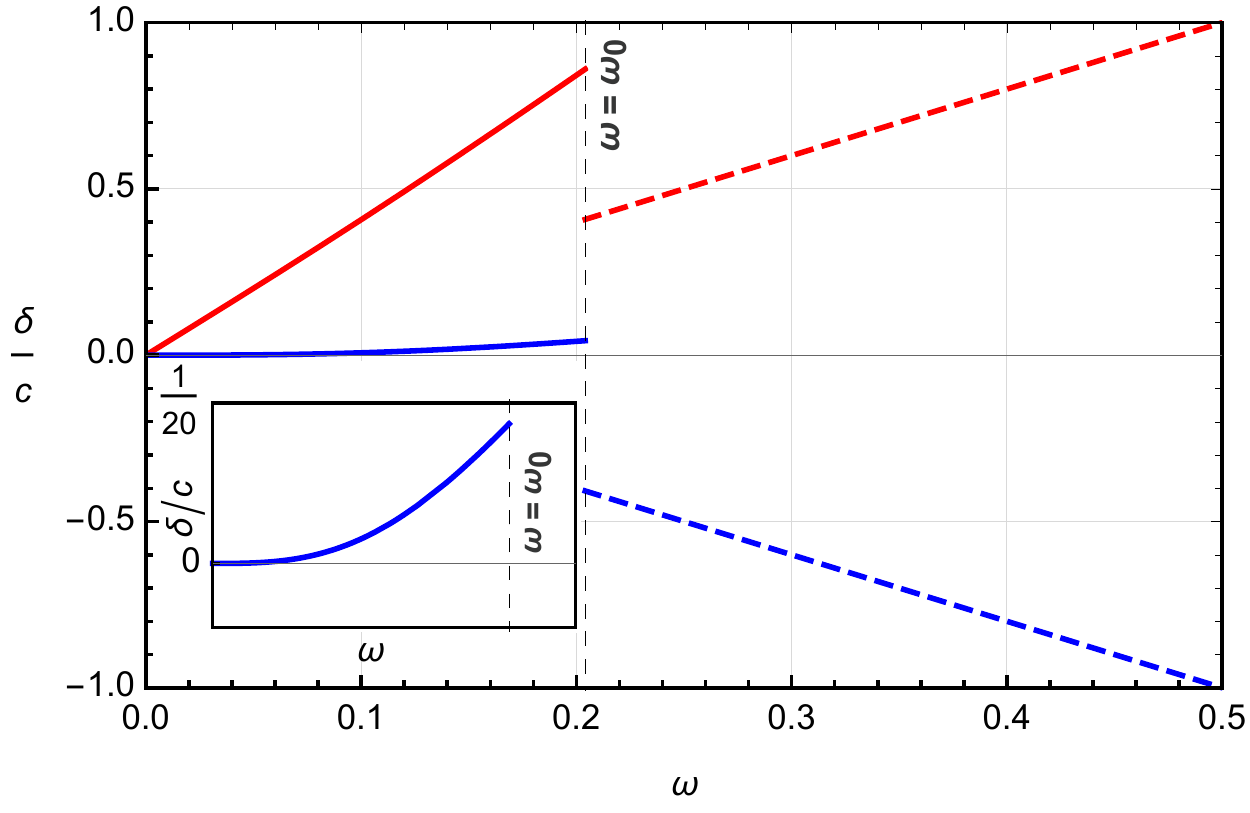}
\par\end{centering}
\caption{\label{plot-rotatory-power-anti-symmetric-case-general} Rotatory power of \eqref{anti-symmetric-case-16} and \eqref{anti-symmetric-case-rotatory-extra-1}. The solid lines indicate the rotatory power (\ref{anti-symmetric-case-rotatory-extra-1}) defined in the region $0<\omega <\omega_{0}$. The dashed curves represent the rotatory power (\ref{anti-symmetric-case-16}) defined in the region $\omega >\omega_{0}$. The vertical dashed line is given by $\omega_{0}= 1/(2\sqrt{6})$, with $\omega_{0}$ of \eqref{anti-symmetric-case-extra-1}. Here we have used $\mu=1$, $\epsilon=2$, $b=1$, $\cos^{2}\theta=1$ and $\alpha''=2$ (red lines) and $\alpha''=-2$ (blue curves). The inset plot highlights the behavior of $\delta$ (\ref{anti-symmetric-case-rotatory-extra-1}) in the regime $0 <\omega < \omega_{0}$.}
\end{figure}

The behavior of the rotatory power for all frequency domain is obtained by plotting \eqref{anti-symmetric-case-rotatory-extra-1} at $0 <\omega <\omega_{0}$ and \eqref{anti-symmetric-case-16} at $\omega >\omega_{0}$, 
with a discontinuity at $\omega=\omega_{0}$, as illustrated in Fig.~\ref{plot-rotatory-power-anti-symmetric-case-general}. In more details, we highlight:
\begin{itemize}
\item for $\alpha''>0$, the rotatory power is always positive, with a discontinuity at $\omega=\omega_{0}$, indicating clockwise rotation of a linear polarization wave;

\item for $\alpha''<0$, the rotatory power is positive for $0 <\omega <\omega_{0}$, and negative for $\omega >\omega_{0}$, with discontinuity and sign reversal at the frequency $\omega=\omega_{0}$.
\end{itemize}

The RP (\ref{anti-symmetric-case-rotatory-extra-1}) exhibits a sign reversal (when $\alpha''<0$) and also discontinuity at the frequency $\omega=\omega_{0}$. This happens because $\delta$ assumes different functional forms for the two frequency intervals under examination. Indeed, for $0<\omega <\omega_{0}$,  $\delta$ is defined by \eqref{anti-symmetric-case-rotatory-extra-1}, while for $\omega >\omega_{0}$ it is given by \eqref{anti-symmetric-case-16}.

\begin{figure}[h!]
\begin{centering}
\includegraphics[scale=0.68]{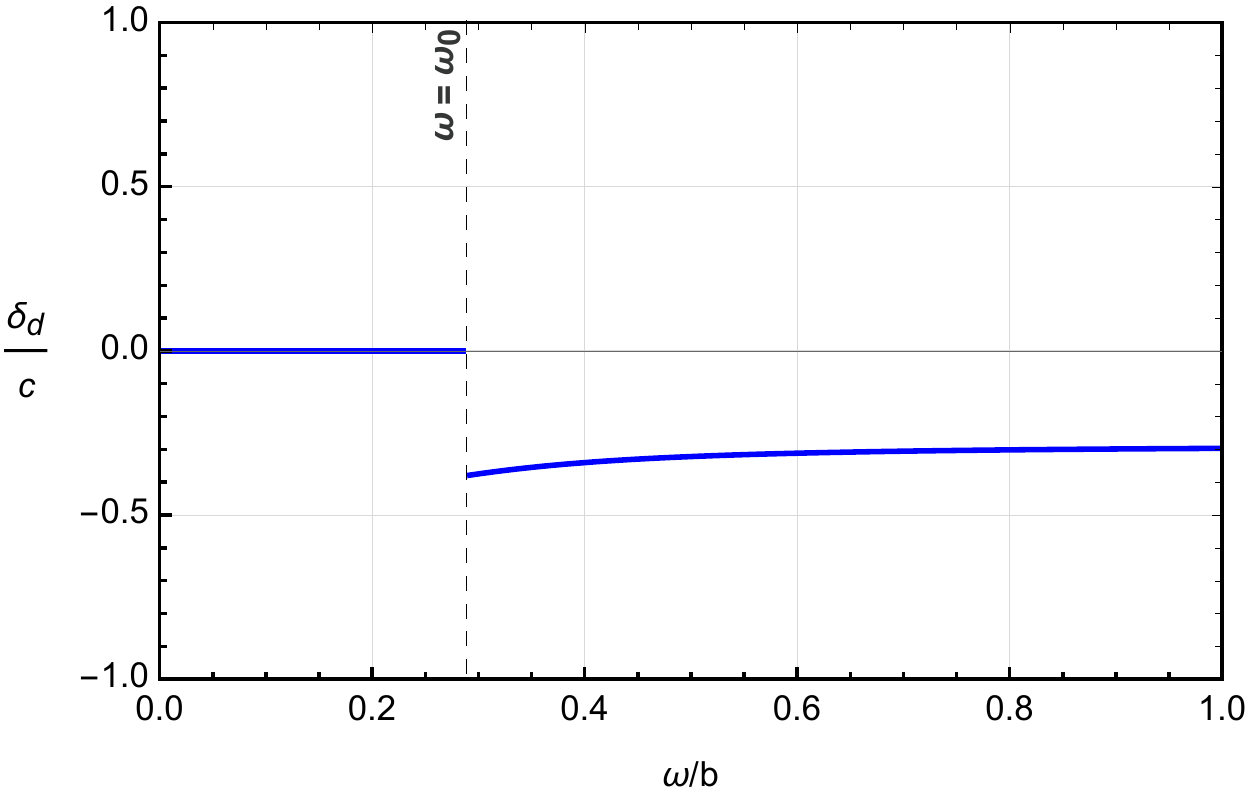}
\par\end{centering} 
\caption{\small{Dichroism coefficient of \eqref{anti-symmetric-case-17} in terms of $\omega/b$. Here, we have used $\mu=1$, $\epsilon=2$, $\alpha''=1$, and $\cos\theta=\pm 1$. \label{plot-dichroism-factor-anti-symmetric}}}
\end{figure}

Another relevant point is that the modes are absorbed to a different degree when $\omega >\omega_{0}$. This difference is characterized by the dichroism coefficient, defined as 
\begin{align}
\delta_{d} &= - \frac{\omega}{2} \left[ \mathrm{Im}(n_{+}) - \mathrm{Im}(n_{-}) \right] , \label{anti-symmetric-case-17}
\end{align}
which for \eqref{anti-symmetric-case-7-0}, yields
\begin{align}
\delta_{d} &=-\omega U \sqrt{\sqrt{1+ \left( \frac{ \mu^{2} c^{2} \alpha'' b\cos\theta}{2 \omega U^{2}} \right)^{2}}-1} . \label{anti-symmetric-case-17}
\end{align} 
The behavior of $\delta_{d}$ for $\cos\theta=\pm1$ in terms of the parameter $\omega /b$ is depicted in Fig. \ref{plot-dichroism-factor-anti-symmetric}, which reveals absence of dichroism for $\omega<\omega_{0}$ and LCR mode, ${\bf{E}}_{+}$, being more absorbed than the RCP mode, ${\bf{E}}_{-}$, for $\omega >\omega_{0}$, that is, $\delta_{d} <0$. Furthermore, it does not occur $\delta_{d}$ sign reversal.


\section{\label{final-remarks}Final Remarks}

In this work, we have examined the classical electromagnetic propagation in bi-isotropic media supporting magnetic current associated with the CME, describing optical effects of the medium. For the full isotropic case of Sec.~\ref{full-isotropic-case}, defined by bi-isotropic constitutive relations and isotropic magnetic conductivity, we have found circularly polarized propagating modes, yielding circular birefringence expressed in terms of the linear dispersive RP (\ref{isotropic-case-1-4}). It experiences sign reversal at the frequency $\omega=\omega'$, with $\omega'$ given by \eqref{cutoff-frequency-full-isotropic-case}.  Such an effect only occurs when $\Sigma \neq 0$ and $\alpha''<0$.
In the scenario of Sec.~\ref{bi-isotropic-anti-symmetric}, the bi-isotropic dielectric is endowed with an asymmetric magnetic current represented by the antisymmetric conductivity tensor parametrized in terms of the constant 3-vector, ${\bf{b}}$, as pointed out in \eqref{anti-symmetic-constitutive-parameters}. The propagating modes are composed of a transversal sector, described by circular polarizations, and a longitudinal component. For the  ${\bf{b}}$-longitudinal special case, the electric fields recover left- and right-handed circularly polarized vectors. The birefringence is examined in terms of the RP considering two frequency intervals: i) $0 <\omega < \omega_{0}$ and ii) $\omega >\omega_{0}$, with $\omega_{0}$ of \eqref{anti-symmetric-case-extra-1}. We have obtained an involved dispersive RP, which exhibits a discontinuity at $\omega = \omega_{0}$ and changes sign whenever $\alpha''<0$. RP reversal is not observed in usual dielectrics, being reported in rotating plasmas \cite{Gueroult} and graphene systems \cite{Poumirol}. The unusual feature here obtained may provide a channel to characterize chiral media. Indeed, it may be used to distinguish the bi-isotropic scenario with antisymmetric conductivity, described by constitutive relations (\ref{anti-symmetic-constitutive-parameters}), from the bi-isotropic medium endowed with isotropic magnetic conductivity (\ref{isot5}).

When the rotatory power, $\delta$, is positive (negative), the medium is defined as right- (left-) handed, since it rotates the plane of linear polarization light in the clockwise (counterclockwise) direction \cite{Wagniere, Hecht}. This definition characterizes the handedness of the medium considering the optical rotation of the linear polarization wave. Thus, the rotatory power inversion here reported also reveals a handedness reversal of the bi-isotropic dielectric (with $\alpha''<0$) under the presence of the magnetic current. Concerning the results of Sec. \ref{full-isotropic-case}, as illustrated in Fig. (\ref{plot-rotatory-dispersion-fullisotropic-case}), blue line, for $\omega <\omega'$ the medium is right-handed, while for $\omega >\omega'$ it becomes left-handed. In the antisymmetric case of Sec.~\ref{bi-isotropic-anti-symmetric}, the Fig. (\ref{plot-rotatory-power-anti-symmetric-case-general}) (blue line) shows a right-handed medium for $0<\omega <\omega_{0}$, and a left-handed medium for $\omega >\omega_{0}$. This can open an interesting connection between electromagnetic chirality/helicity \cite{TangPRL, Cameron,Bliokh, Guasti,Qiu} and the combined constitutive relations examined in this work.

\subsection*{Acknowledgments}

The authors P.D.S.S. and M.M.F.Jr. express their gratitude to FAPEMA, CNPq, and CAPES (Brazilian research agencies) for their invaluable financial support. M.M. F. is supported by FAPEMA Universal/01187/18,  CNPq/Produtividade 311220/2019-3 and CNPq/Universal/422527/2021-1.
Furthermore, we are indebted to CAPES/Finance Code 001 and FAPEMA/POS-GRAD-02575/21.


\begin{thebibliography}{99}
\bibitem{Jackson} J.D. Jackson, \textit{Classical Electrodynamics}, 3rd ed.
(John Wiley \& Sons, New York, 1999).

\bibitem{Zangwill} A.~Zangwill, \textit{Modern Electrodynamics} (Cambridge
University Press, New York, 2012).








\bibitem{Sihvola} I. V. Lindell, A. H. Sihvola, S. A. Tretyakov, and A. J. Viitanen,
\textit{Electromagnetic Waves in Chiral and Bi-Isotropic Media} (Artech House, Boston, 1993).

\bibitem{Kong} J. A. Kong, \textit{Electromagnetic Wave Theory} (Wiley, New York, 1986).








\bibitem{Nieves} J. F. Nieves and P. B. Pal, Third electromagnetic constant
of an isotropic medium, \href{http://dx.doi.org/10.1119/1.17598}{Am. J.
Phys. \textbf{62}, 207 (1994)}.

\bibitem{Gauthier} R. C. Gauthier, Bi-anisotropic resonators analyzed using
Fourier--Bessel numerical formulation; Sagnac effect application, \href{https://doi.org/10.1016/j.optcom.2018.11.026}{Optics Communications 435, 413 (2019)}.

\bibitem{Aladadi} Y. T. Aladadi and M. A. S. Alkanhal, Classification and
characterization of electromagnetic materials, \href{https://doi.org/10.1038/s41598-020-68298-3}%
{Sci. Rep. \textbf{10}, 11406 (2020)}.



\bibitem{Sihvola1} A. H. Sihvola and I. V. Lindell, Bi-isotropic
constitutive relations, \href{https://doi.org/10.1002/mop.4650040805}{%
Microw. Opt. Technol. Lett., \textbf{4} (8), 295-297 (1991)}.

\bibitem{Sihvola2} A. H. Sihvola and I. V. Lindell, Properties of
bi-isotropic Fresnel reflection coefficients, \href{https://www.sciencedirect.com/science/article/abs/pii/003040189290237L?via\%3Dihub}{Optics Communications 89, 1
(1992)}.











\bibitem{Chang1} Ming-Che Chang and Min-Fong Yang, Optical signature of
topological insulators, \href{https:10.1103/PhysRevB.80.113304}{Phys. Rev. B
\textbf{80}, 113304 (2009)}.

\bibitem{Urrutia} A. Mart\'in-Ruiz, M. Cambiaso, and L. F. Urrutia, The
magnetoelectric coupling in Electrodynamics. \href{https://www.worldscientific.com/doi/abs/10.1142/S0217751X19410021}%
{ Int. J. Mod. Phys. A \textbf{34}, 1941002 (2019)}.

\bibitem{Urrutia2} A.~Mart\'{\i}n-Ruiz, M.~Cambiaso, and L.F.~Urrutia,
Electro- and magnetostatics of topological insulators as modeled by planar,
spherical, and cylindrical $\theta$ boundaries: ForestGreen's function
approach, \href{https://journals.aps.org/prd/abstract/10.1103/PhysRevD.93.045022}%
{Phys. Rev. D \textbf{93}, 045022 (2016)}.

\bibitem{Lakhtakia} A. Lakhtakia and T. G. Mackay, Classical electromagnetic
model of surface states in topological insulators, \href{https://doi.org/10.1117/1.JNP.10.033004}%
{J. Nanophoton. \textbf{10} (3), 033004 (2016)}.

\bibitem{Winder} T. M. Melo, D. R. Viana, W. A. Moura-Melo, J. M. Fonseca,
A. R. Pereira, Topological cutoff frequency in a slab waveguide: Penetration
length in topological insulator walls, \href{https://doi.org/10.1016/j.physleta.2015.12.041}%
{Phys, Lett. A \textbf{380}, 973 (2016).}

\bibitem{Li} Z.-X. Li, Yunshan Cao, Peng Yan, Topological insulators and
semimetals in classical magnetic systems, \href{https://doi.org/10.1016/j.physrep.2021.02.003}%
{Phys. Report \textbf{915}, 1 (2021).}

\bibitem{Li1} R. Li, J. Wang, Xiao-Liang Qi and S.-C. Zhang, Dynamical axion
field in topological magnetic insulators, \href{https://doi.org/10.1038/NPHYS1534}%
{Nature Phys. \textbf{6}, 284 (2010)}.

\bibitem{Sekine} A. Sekine and K. Nomura, Axion electrodynamics in
topological materials, \href{https://doi.org/10.1063/5.0038804}{J. Appl.
Phys. \textbf{129}, 141101 (2021).}

\bibitem{Tobar} M. E. Tobar, B. T. McAllister, and M. Goryachev, Modified
axion electrodynamics as impressed electromagnetic sources through
oscillating background polarization and magnetization, \href{https://doi.org/10.1016/j.dark.2019.100339}%
{Phys. Dark Universe \textbf{26}, 100339 (2019)}.

\bibitem{BorgesAxion} L. H. C. Borges, A. G. Dias, A. F. Ferrari, J. R.
Nascimento, A. Yu. Petrov, Generation of Axion-Like Couplings via Quantum
Corrections in a Lorentz Violating Background, \href{https://doi.org/10.1103/PhysRevD.89.045005}%
{ Phys. Rev. D \textbf{89}, 045005 (2014)}.





\bibitem{Lorenci} V. A. De Lorenci and G. P. Goulart, Magnetoelectric birefringence revisited, \href{https://doi.org/10.1103/PhysRevD.78.045015} {Phys. Rev. D {\bf{78}}, 045015 (2008)}.








\bibitem{Helayel1}
J. M. A. Paix\~ao, L. P. R. Ospedal, M. J. Neves, and J. A. Helay\"el-Neto, The axion-photon mixing in non-linear electrodynamic scenarios, \href{https://doi.org/10.48550/arXiv.2205.05442}{arXiv:2205.05442 (2022)}.





\bibitem{Helayel2}
P. Gaete and J. A. Helayel-Neto, Vacuum material properties and Cherenkov radiation in Logarithmic Electrodynamics, \href{https://doi.org/10.48550/arXiv.2205.03252}{arXiv:2205.03252 (2022)}.



\bibitem{Yakov} I. Yakov, Dispersion relation for electromagnetic waves in anisotropic media, \href{https://doi.org/10.1016/j.physleta.2009.12.071}{Phys. Lett. A \textbf{374}, 1113 (2010)}.
\bibitem{Damaskos} N.J. Damaskos, A.L. Maffett and P.L.E. Uslenghi, Dispersion relation for general anisotropic media, \href{https://ieeexplore.ieee.org/document/1142905}{IEEE Trans. Antennas Propagat. AP-30, 991 (1982)}.



\bibitem{Lin} Shi-Rong Lin, Ruo-Yang Zhang, Yi-Rong Ma, W. Jia, Q. Zhaoa, 
Electromagnetic wave propagation in time-dependent media with antisymmetric magnetoelectric coupling,
\href{http://dx.doi.org/10.1016/j.physleta.2016.05.050} {Phys. Lett. A \textbf{380}, 2582 (2016)}.


\bibitem{Sihvola3} S. Ougier, I. Chenerie, A. Sihvola, and A. Priou,
Propagation in bi-isotropic media: effect of different formalisms on the
propagation analysis, \href{http://www.jpier.org/PIER/pier.php?paper=9301010}%
{Progress In Electromagnetics Research \textbf{09}, 19 (1994)}.

\bibitem{Pedro3}
P. D. S.~Silva, R. Casana, and M. M.~Ferreira Jr., Symmetric and antisymmetric constitutive tensors for bi-isotropic and bi-anisotropic media, \href{https://doi.org/10.48550/arXiv.2204.10460} {arXiv:2204.10460 (2022)}.

\bibitem{Hillion} P. Hillion, Manifestly covariant formalism for
electromagnetism in chiral media, \href{https://doi.org/10.1103/PhysRevE.47.1365}%
{Phys. Rev. E \textbf{47}, 1365 (1993)}.

\bibitem{Kaushik2}
S. Kaushik, Magnetic and Optical Response of Chiral Fermions, (Ph.D. thesis), \href{https://arxiv.org/abs/2112.13749}{arXiv:2112.13749 (2021)}.

\bibitem{Barron2} L. D. Barron, \textit{Molecular Light Scattering and Optical Activity}, 2nd ed. (Cambridge University Press, New York, 2004).

 \bibitem{Hecht} E. Hecht, \textit{Optics}, 4nd ed. (Addison Wesley, San
Francisco, 2002).

\bibitem{Wagniere}  G. H. Wagniere, On Chirality and the Universal Asymmetry: Reflections on Image and Mirror Image,  ‎(Wiley-Vch, Zurich) (2007).

 \bibitem{TangPRL} Y. Tang and A. E. Cohen,  Optical Chirality and Its Interaction with Matter, \href{https://doi.org/10.1103/PhysRevLett.104.163901}{Phys. Rev. Lett. {\bf{104}}, 163901 (2010)}.
 
 \bibitem{Fowles} G. R. Fowles, \textit{Introduction to modern optics}, 2nd
ed. (Dover Publications, INC., New York, 1975); A. K. Bain, \textit{Crystal optics: properties and applications} (Wiley-VCH Verlag GmbH \& Co. KGaA, Germany, 2019).


\bibitem{Condon} E. U. Condon, Theories of Optical Rotatory Power, \href{https://doi.org/10.1103/RevModPhys.9.432}{Rev. Mod. Phys. {\bf{9}}, 432 (1937)}.

\bibitem{Bennett}H. S. Bennett, E. A. Stern, Faraday effect in solids. \href{https://doi.org/10.1103/PhysRev.137.A448}{Phys. Rev. \textbf{137}, A448--A461 (1965)}; L. M. Roth. Theory of the Faraday effect in solids, \href{https://doi.org/10.1103/PhysRev.133.A542}{Phys. Rev. \textbf{133}, A542--A553 (1964)}.

\bibitem{Ohnoutek} L. Ohnoutek \textit{et. al.}, Strong interband Faraday rotation in 3D topological insulator $\mathrm{Bi}_{2}\mathrm{Se}_{3}$, \href{https://doi.org/10.1038/srep19087} {Sci. Rep.\textbf{6}, 19087 (2016).}


\bibitem{Tse}W.-K. Tse, A. H. MacDonald, Giant magneto-optical Kerr effect and universal Faraday effect in thin-film topological insulators, \href{https://doi.org/10.1103/PhysRevLett.105.057401}{Phys. Rev. Lett. {\bf{105}}, 057401 (2010)}; Magneto-optical and magnetoelectric effects of topological insulators in quantizing magnetic fields,	 \href{https://doi.org/10.1103/PhysRevB.82.161104} {Phys. Rev. B {\bf{82}}, 161104 (2010)};  Magneto-optical Faraday and Kerr effects in topological insulator films and in other layered quantized Hall systems, \href{https://doi.org/10.1103/PhysRevB.84.205327}{Phys. Rev. B {\bf{84}}, 205327 (2011)}.
	
\bibitem{Maciejko} J. Maciejko, X.-L. Qi, H. D. Drew, S.-C. Zhang. Topological quantization in units of the fine structure constant, \href{https://doi.org/10.1103/PhysRevLett.105.166803}{Phys. Rev. Lett. {\bf{105}}, 166803 (2010)}.

\bibitem{Crasee}I. Crassee, J. Levallois, A. L. Walter, M. Ostler, A. Bostwick, E. Rotenberg, T. Seyller, D. van der Marel, and A. B. Kuzmenko, Giant Faraday rotation in single- and multilayer graphene, \href{https://doi.org/10.1038/nphys1816}{Nature Phys. {\bf{7}}, 48 (2011)}.

\bibitem{Shimano}R. Shimano, G. Yumoto, J. Y. Yoo, R. Matsunaga, S. Tanabe, H. Hibino, T. Morimoto and H. Aoki, Quantum Faraday and Kerr
rotations in graphene, \href{https://doi.org/10.1038/ncomms2866}{Nature Comm. {\bf{4}}, 1841 (2013)}.


\bibitem{Tschugaeff} L. Tschugaeff, Anomalous rotatory dispersion, \href{https://doi.org/10.1039/TF9141000070}{Trans. Faraday Soc., {\bf{10}}, 70-79 (1914)}.

\bibitem{Newnham} R. E. Newnham, \textit{Properties of Materials - anisotropy, symmetry, structure} (Oxford University Press, New York, 2005).

 \bibitem{Dimitriu} D. G. Dimitriu, D. O. Dorohoi, New method to determine the optical rotatory dispersion of inorganic crystals applied to some samples of Carpathian Quartz, \href{https://doi.org/10.1016/j.saa.2014.04.139}{Spectrochimica Acta Part A: Molecular and Biomolecular Spectroscopy {\bf{131}}, 674-677 (2014)}.
 
 \bibitem{Birefringence1} L.A.~Pajdzik and A.M.~Glazer, Three-dimensional birefringence imaging with a microscope tilting-stage. I. Uniaxial
crystals, \href{https://doi.org/10.1107/S0021889806007758}{J. Appl. Cryst. {\bf 39}, 326 (2006)}.
 

 
 

 
\bibitem{Xing-Liu} X. Liu, J. Yang, Z. Geng, nad H. Jia, Simultaneous measurement of optical rotation dispersion and absorption spectra for chiral substances, \href{https://doi.org/10.1002/chir.23233} {Chirality {\bf{8}}, 32, 1071-1079 (2022)}. 








 
 
 
 
 \bibitem{Gueroult} R. Gueroult, J-M. Rax, and N. J. Fisch, Enhanced tuneable rotatory power in a rotating plasma, \href{https://doi.org/10.1103/PhysRevE.102.051202}{Phys. Rev. E {\bf{102}}, 051202 (R), 2020}.
 


\bibitem{Gueroult2} R. Gueroult, Y. Shi, J-M. Rax, and N. J. Fisch, Determining the rotation direction in pulsars, \href{https://doi.org/10.1038/s41467-019-11243-4}{Nat. Commun. {\bf{10}}, 3232 (2019)}.



\bibitem{Poumirol} J-M. Poumirol, P. Q. Liu, T. M. Slipchenko, A. Y. Nikitin, L. Martin-Morento, J. Faist, and A. B. Kuzmenko, Electrically controlled terahertz magneto-optical phenomena in continuous and patterned graphene, \href{https://doi.org/10.1038/ncomms14626} {Nat. Commun. {\bf{8}}, 14626 (2017)}.



\bibitem{Tutunnikov} I. Tutunnikov, U. Steinitz, E. Gershnabel, J-M. Hartmann, A. A. Milner, V. Milner, and I. Sh. Averbukh, Rotation of the polarization of light as a tool for investigating the collisional transfer of angular momentum from rotating molecules to macroscopic gas flows, \href{https://doi.org/10.1103/PhysRevResearch.4.013212} {Phys. Rev. Research {\bf{4}}, 013212 (2022)}.


\bibitem{Steinitz} U. Steinitz and I. Sh. Averbukh, Giant polarization drag in a gas of molecular super-rotors, \href{https://doi.org/10.1103/PhysRevA.101.021404}{ Phys. Rev. A {\bf{101}}, 021404(R) (2020).} 




\bibitem{Tischler} N. Tischler, M. Krenn, R. Fickler, X. Vidal, A. Zeilinger, and G. Molina-Terriza, Quantum optical rotatory dispersion, \href{https://doi.org/10.1126/sciadv.1601306}{Sci. Adv. {\bf{2}}, e1601306 (2016)}.









\bibitem{Woo} J. H. Woo, B. K. M. Gwon, J. H. Lee, D-W. Kim, W. Jo, D. H. Kim, and J. W. Wu, Time-resolved pump-probe measurement of optical rotatory dispersion in chiral metamaterial, \href{https://doi.org/10.1002/adom.201700141} {Adv. Optical Mater.  {\bf{5}}, 1700141 (2017)}.





\bibitem{Zhang} Q. Zhang, E. Plum, J-Y. Ou, H. Pi, J. Li, K. F. MacDonald, and N. I. Zheludev, Electrogyration in metamaterials: chirality and polarization rotatory power that depend on applied electric field, \href{https://doi.org/10.1002/adom.202001826} {Adv. Optical Mater. {\bf{9}}, 2001826 (2021)}.


\bibitem{Hosur}
P. Hosur, and X-L. Qi, Tunable circular dichroism due to the chiral anomaly in Weyl semimetals, \href{https://doi.org/10.1103/PhysRevB.91.081106}{Phys. Rev. B \textbf{91}, 081106(R) (2015)}.


\bibitem{Nieto-Vesperinas}
M. Nieto-Vesperinas, Optical theorem for the conservation of electromagnetic helicity: Significance for molecular energy transfer and enantiomeric discrimination by circular dichroism, \href{https://doi.org/10.1103/PhysRevA.92.023813} {Phys. Rev. A \textbf{92}, 023813 (2015)}.


\bibitem{Tang} Y. Tang and A. E. Cohen, Enhanced Enantioselectivity in Excitation of Chiral Molecules by Superchiral Light,   \href{https://doi.org/10.1126/science.1202817} {Science \textbf{332}, 333 (2011)}.


\bibitem{Amin}
M. Amin, O. Siddiqui, and M. Farhat, Linear and circular dichroism in graphene-based reflectors for polarization control, \href{https://doi.org/10.1103/PhysRevApplied.13.024046}{Phys. Rev. Applied \textbf{13}, 024046 (2020)}.






\bibitem{Kharzeev1} D.E.~Kharzeev, The chiral magnetic effect and anomaly-induced transport, \href{https://doi.org/10.1016/j.ppnp.2014.01.002}{Prog. Part. Nucl. Phys. {\bf 75}, 133 (2014)};
D.E.~Kharzeev, J.~Liao, S.A.~Voloshin, and G.~Wang, Chiral magnetic and vortical effects in high-energy nuclear collisions -- A status report, \href{https://doi.org/10.1016/j.ppnp.2016.01.001}{Prog. Part. Nucl. Phys. \textbf{88}, 1 (2016)};
\bibitem{Kharzeev1B}D.~Kharzeev, K.~Landsteiner, A.~Schmitt and H.U.~Yee, \textit{Strongly Interacting Matter in Magnetic Fields},
Lect. Notes Phys. \textbf{871} (Springer-Verlag, Berlin $\cdot$ Heidelberg, 2013).

\bibitem{Fukushima} K.~Fukushima, D.E.~Kharzeev, and H.J.~Warringa, Chiral magnetic effect, \href{https://doi.org/10.1103/PhysRevD.78.074033}{Phys. Rev. D \textbf{78}, 074033 (2008)}.

\bibitem{Gabriele} G.~Inghirami, M.~Mace, Y.~Hirono, L.~Del Zanna, D.E.~Kharzeev, and M.~Bleicher, Magnetic fields in heavy ion collisions: flow and charge transport,  \href{https://link.springer.com/article/10.1140/epjc/s10052-020-7847-4}{Eur. Phys. J. C \textbf{80}, 293 (2020)}.


\bibitem{Bubnov} A.F.~Bubnov, N.V.~Gubina, and V.Ch.~Zhukovsky, Vacuum current induced by an axial-vector condensate and electron anomalous magnetic moment in a magnetic field, \href{https://doi.org/10.1103/PhysRevD.96.016011}{Phys. Rev. D \textbf{96}, 016011 (2017)}.



\bibitem{Burkov} A.A.~Burkov, Chiral anomaly and transport in Weyl metals, \href{https://doi.org/10.1088/0953-8984/27/11/113201}{J. Phys. Condens. Matter \textbf{27}, 113201 (2015)}.



\bibitem{Chang} M.-C.~Chang and M.-F.~Yang, Chiral magnetic effect in a two-band lattice model of Weyl semimetal, \href{https://doi.org/10.1103/PhysRevB.91.115203}{Phys. Rev. B \textbf{91}, 115203 (2015)}.


\bibitem{Wurff} E.C.I~van der Wurff and H.T.C.~Stoof, Anisotropic chiral magnetic effect from tilted Weyl cones, \href{https://doi.org/10.1103/PhysRevB.96.121116}{Phys. Rev. B \textbf{96}, 121116(R) (2017)}.


\bibitem{Landsteiner} K.~Landsteiner, Anomalous transport of Weyl fermions in Weyl semimetals, \href{https://doi.org/10.1103/PhysRevB.89.075124}{Phys. Rev. B \textbf{89}, 075124 (2014)}.


\bibitem{Kaushik} S.~Kaushik and D.E.~Kharzeev, Quantum oscillations in the chiral magnetic conductivity, \href{https://doi.org/10.1103/PhysRevB.95.235136}{Phys. Rev. B \textbf{95}, 235136 (2017)}.


\bibitem{Ruiz} A.~Mart\'{i}n-Ruiz, M.~Cambiaso, and L.F.~Urrutia, Electromagnetic fields induced by an electric charge near a Weyl semimetal, \href{https://doi.org/10.1103/PhysRevB.99.155142}{Phys. Rev. B \textbf{99}, 155142 (2019)}.

\bibitem{Pedro1} P.D.S.~Silva, M.M.~Ferreira Jr., M.~Schreck, and L.F.~Urrutia, Magnetic-conductivity effects on electromagnetic propagation in
dispersive matter, \href{https://journals.aps.org/prd/abstract/10.1103/PhysRevD.102.076001}{Phys. Rev. D {\bf{102}}, 076001 (2020)}.



\bibitem{Kaushik1} S.~Kaushik, D.E.~Kharzeev, and E.J.~Philip, Transverse chiral magnetic photocurrent induced by linearly polarized light in symmetric Weyl semimetals, \href{https://doi.org/10.1103/PhysRevResearch.2.042011}{Phys. Rev. Research {\bf{2}}, 042011(R) (2020)}.


\bibitem{Pedro2}
P.D.S.~Silva, L. Lisboa-Santos, M. M.~Ferreira Jr., and M.~Schreck, Effects of CPT-odd terms of dimensions three and five on electromagnetic propagation in continuous matter, \href{https://doi.org/10.1103/PhysRevD.104.116023} {Phys. Rev. D {\bf{104}}, 116023 (2021)}.



\bibitem{Marco}A Kostelecky, R. Lehnert, N. McGinnis, M. Schreck, B. Seradjeh, Lorentz violation in Dirac and Weyl semimetals, 
\href{https://doi.org/10.48550/arXiv.2112.14293}{Phys. Rev. Research {\bf{4}}, 023106 (2022)}.




\bibitem{Mun} J. Mun, et al. Electromagnetic chirality: from fundamentals to nontraditional chiroptical phenomena. \href{https://doi.org/10.1038/s41377-020-00367-8} {Light Sci. Appl. {\bf{9}}, 139 (2020).}


\bibitem{Cameron}R. P. Cameron, S. M. Barnett and A. M. Yao,  Optical helicity, optical spin and related quantities in electromagnetic theory, \href{https://doi.org/10.1088/1367-2630/14/5/053050} {New J. Phys. {\bf{14}} 053050 (2012)}.

\bibitem{Bliokh}K. Y Bliokh, A. Y. Bekshaev and F. Nori, Dual electromagnetism: helicity, spin, momentum and angular momentum, \href{https://doi.org/10.1088/1367-2630/15/3/033026} {New J. Phys. {\bf{15}} 033026 (2013)}; Corrigendum: Dual electromagnetism: helicity, spin, momentum,
and angular momentum, \href{https://doi.org/10.1088/1367-2630/18/8/089503} {New J. Phys. {\bf{18}} 089503 (2016)}; Conservation of the spin and orbital angular momenta
in electromagnetism, \href{https://doi.org/10.1088/1367-2630/16/9/093037} {New J. Phys. 116  093037 (2014)}.

\bibitem{Guasti} M.F. Guasti, Chirality, helicity and the rotational content of electromagnetic fields, \href{https://doi.org/10.1016/j.physleta.2019.06.002}{Phys. Lett. A {\bf{383}}, 3180 (2019)}.


\bibitem{Qiu}C.-W. Qiu, H.-Y. Yao, L.-W. Li, S. Zouhdi, and T.-S. Yeo,  Routes to left-handed materials by magnetoelectric couplings, \href{10.1103/PhysRevB.75.245214} {Phys. Rev. B {\bf{75}}, 245214 (2007)}.

\end{thebibliography}
\end{document}